\documentclass[onecolumn,showpacs]{revtex4}

\usepackage{graphicx}% Include figure files
\usepackage{dcolumn}% Align table columns on decimal point
\usepackage{bm}% bold math

\input epsf

\begin{document}

\title{\Large Is Modified Chaplygin gas along with barotropic fluid responsible for acceleration of the Universe?}

\author{\bf Writambhara Chakraborty$^1$\footnote{writam1@yahoo.co.in} and
Ujjal Debnath$^2$\footnote{ujjaldebnath@yahoo.com}}

\affiliation{$^1$Department of Mathematics, Heritage Institute of
Technology, Anandapur, Kolkata-700 107, India.\\
$^2$Department of Mathematics, Bengal Engineering and Science
University, Shibpur, Howrah-711 103, India.}

\date{\today}

\begin{abstract}
In this letter, we have considered a model of the universe filled
with modified Chaplygin gas and another fluid (with barotropic
equation of state) and its role in accelerating phase of the
universe. We have assumed that the mixture of these two fluid
models is valid from (i) the radiation era to $\Lambda$CDM for
$-1\le\gamma\le 1$  and (ii) the radiation era to quiessence model
for $\gamma<-1$. For these two fluid models, the statefinder
parameters describe different phase of the evolution of the
universe.
\end{abstract}


\maketitle

Recent measurements of redshift and luminosity-distance relations
of type Ia Supernovae indicate that the expansion of the Universe
is accelerating [1-5]. This implies that the pressure $p$ and the
energy density $\rho$ of the Universe should violate the strong
energy condition $\rho+3p<0$ i.e., pressure must be negative. The
matter responsible for this condition to be satisfied at some
stage of evolution of the universe is referred to as {\it dark
energy} [6 - 8]. There are different candidates to play the role
of the dark energy. The most traditional candidate is a
non-vanishing cosmological constant which can also be though of
as a perfect fluid satisfying the equation of state $p=-\rho$.
Negative pressure leading to an accelerating Universe can also be
obtained in a Chaplygin gas cosmology [9], in which the matter is
taken to be a perfect fluid obeying an exotic equation of state
$p=-B/\rho, (B>0)$. The Chaplygin gas behaves as pressureless
fluid for small values of the scale factor and as a cosmological
constant for large values of the scale factor which tends to
accelerate the expansion. Subsequently the above equation was
generalized to the form $p=-B/\rho^{\alpha}, 0\le \alpha \le 1$
[10-12] and recently it was modified to the form
$p=A\rho-B/\rho^{\alpha}, (A>0)$ [13, 14], which is known as {\it
Modified Chaplygin Gas}. This equation of state shows a radiation
era ($A=1/3$) at one extreme and a $\Lambda CDM$ model
at the other extreme.\\

The metric of a homogeneous and isotropic universe in FRW model is

\begin{equation}
ds^{2}=dt^{2}-a^{2}(t)\left[\frac{dr^{2}}{1-kr^{2}}+r^{2}(d\theta^{2}+sin^{2}\theta
d\phi^{2})\right]
\end{equation}

where $a(t)$ is the scale factor and $k~(=0,\pm 1)$ is the
curvature scalar.\\

The Einstein field equations are (choosing $8\pi G=c=1$)

\begin{equation}
\frac{\dot{a}^{2}}{a^{2}}+\frac{k}{a^{2}}=\frac{1}{3}\rho
\end{equation}
and
\begin{equation}
\frac{\ddot{a}}{a}=-\frac{1}{6}(\rho+3p)
\end{equation}

The energy conservation equation ($T_{\mu;\nu}^{\nu}=0$) is

\begin{equation}
\dot{\rho}+3\frac{\dot{a}}{a}(\rho+p)=0
\end{equation}

For modified Chaplygin gas, equation (4) yields

\begin{equation}
\rho=\left[\frac{B}{1+A}+\frac{C}{a^{3(1+A)(1+\alpha)}}
\right]^{\frac{1}{1+\alpha}}
\end{equation}

where $C$ is an arbitrary integration constant.\\

Here we consider two fluid cosmological model which besides a
modified Chaplygin's component, with equation of state (4)
contains also a barotropic fluid component with equation of state
$p_{_{1}}=\gamma\rho_{_{1}}$. Normally for accelerating universe
$\gamma$ satisfies $-1\le\gamma\le 1$. But observations state that
$\gamma$ satisfies $-1.6\le\gamma\le 1$, i.e., $\gamma<-1$
corresponds to phantom model. For these two component fluids,
r.h.s of equations (2) and (3), i.e., $\rho$ and $p$ should be
replaced by $\rho+\rho_{_{1}}$ and $p+p_{_{1}}$ respectively. Here
we have assumed the two fluid are separately conserved. For
Chaplygin gas, the density has the expression given in equation
(5) and for another fluid, the conservation equation gives the
expression for density as
\begin{equation}
\rho_{_{1}}=\frac{d}{a^{3(1+\gamma)}}
\end{equation}

where $d$ is an integration constant.\\

We have described this two fluid cosmological model from the field
theoretical point of view by introducing a scalar field $\phi$
and a self-interacting potential $V(\phi)$ with the effective
Lagrangian

\begin{equation}
{\cal L_{\phi}}=\frac{1}{2}\dot{\phi}^{2}-V(\phi)
\end{equation}

The analogous energy density $\rho_{\phi}$ and pressure
$p_{\phi}$ corresponding scalar field $\phi$ having a
self-interacting potential $V(\phi)$ are the following:

\begin{equation}
\rho_{\phi}=\frac{1}{2}~\dot{\phi}^{2}+V(\phi)=\rho+\rho_{_{1}}=\left[\frac{B}{1+A}+\frac{C}{a^{3(1+A)(1+\alpha)}}
\right]^{\frac{1}{1+\alpha}}+\frac{d}{a^{3(1+\gamma)}}
\end{equation}
and
\begin{eqnarray*}
p_{\phi}=\frac{1}{2}~\dot{\phi}^{2}-V(\phi)=p+p_{_{1}}=A\rho-\frac{B}{\rho^{\alpha}}+\gamma\rho_{_{1}}
=A \left[\frac{B}{1+A}+\frac{C}{a^{3(1+A)(1+\alpha)}}
\right]^{\frac{1}{1+\alpha}}
\end{eqnarray*}
\begin{equation}
\hspace{3in} -B \left[\frac{B}{1+A}+\frac{C}{a^{3(1+A)(1+\alpha)}}
\right]^{-\frac{\alpha}{1+\alpha}}+\frac{\gamma~d}{a^{3(1+\gamma)}}
\end{equation}
\\

For flat Universe ($k=0$) and by the choice $\gamma=A$, we have
the expression for $\phi$ and $V(\phi)$:

\begin{equation}
\phi=-\frac{1}{\sqrt{3(1+A)}~(1+\alpha)}\int
\left[\frac{d+c\left(C+\frac{Bz}{1+A}
\right)^{-\frac{\alpha}{1+\alpha}}} {d+\left(C+\frac{Bz}{1+A}
\right)^{\frac{1}{1+\alpha}}} \right]^{\frac{1}{2}}~\frac{dz}{z}
\end{equation}
and
\begin{equation}
V(\phi)=A \left[\frac{B}{1+A}+\frac{C}{z}
\right]^{\frac{1}{1+\alpha}}-B \left[\frac{B}{1+A}+\frac{C}{z}
\right]^{-\frac{\alpha}{1+\alpha}}+\frac{(1-A)~d}{z^{\frac{1}{1+A}}}
\end{equation}

where $z=a^{3(1+A)(1+\alpha)}$.\\

Here we have considered $\gamma=A$ for simplicity. Although by
this choice we consider the two fluids to coincide at high
densities, i.e., for early Universe. Taking $\gamma=A=0$ we
consider the mixture of dust with the generalized Chaplygin gas
which has been discussed in the works of Gorini et al [10, 11].
Also for the choice $\gamma=A=\frac{1}{3}$ both the fluids
represent
radiation as the density is high at radiation era.\\

The graphical representation of $\phi$ against $a$ and $V(\phi)$
against $a$ and $\phi$ respectively have been shown in figures 1 -
3 for $A=1/3$ and $\alpha=1$. From figure 1 we have seen that
scalar field $\phi$ decreases when scale factor $a(t)$ increases
for $A=1/3$. In figure 2, we see that potential function $V(\phi)$
sharply decreases from extremely large value to a fixed value for
$A=1/3$. The potential function $V(\phi)$ increases to infinitely
large value when scale factor $a(t)$ increases for $A=1/3$. So the
figures show how $V(\phi)$ varies with $\phi$ and $a(t)$.\\

\begin{figure}
\includegraphics[height=1.7in]{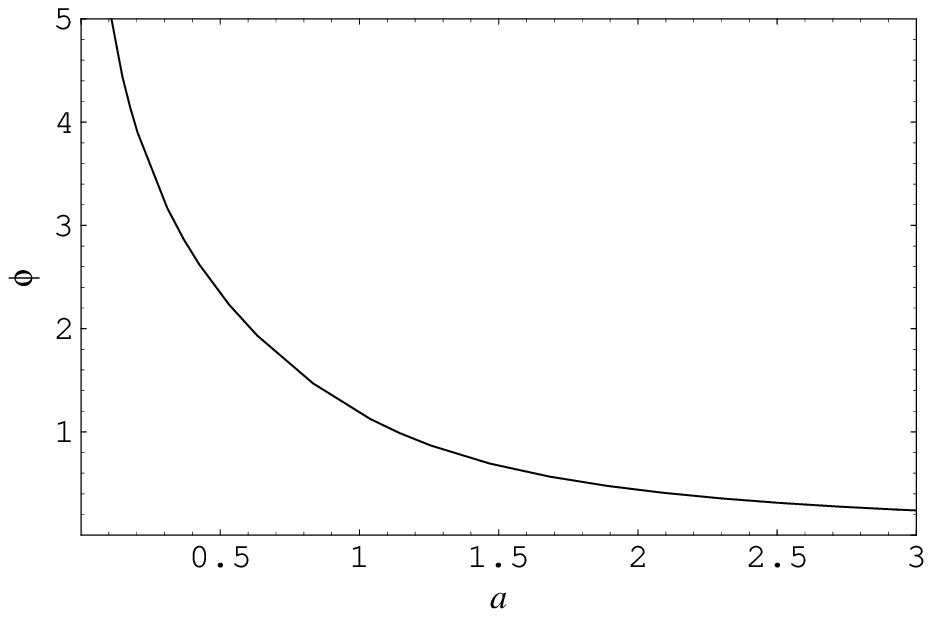}~~~
\includegraphics[height=1.7in]{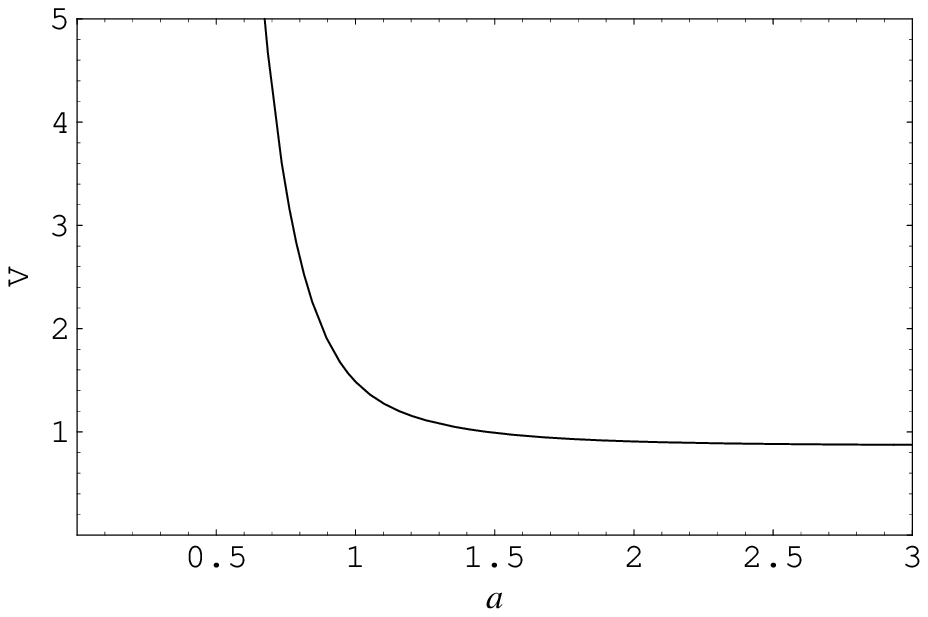}\\
\vspace{1mm}
Fig.1~~~~~~~~~~~~~~~~~~~~~~~~~~~~~~~~~~~~~~~~~~~~~~~~~~~~~Fig.2\\
\vspace{5mm}
\includegraphics[height=1.7in]{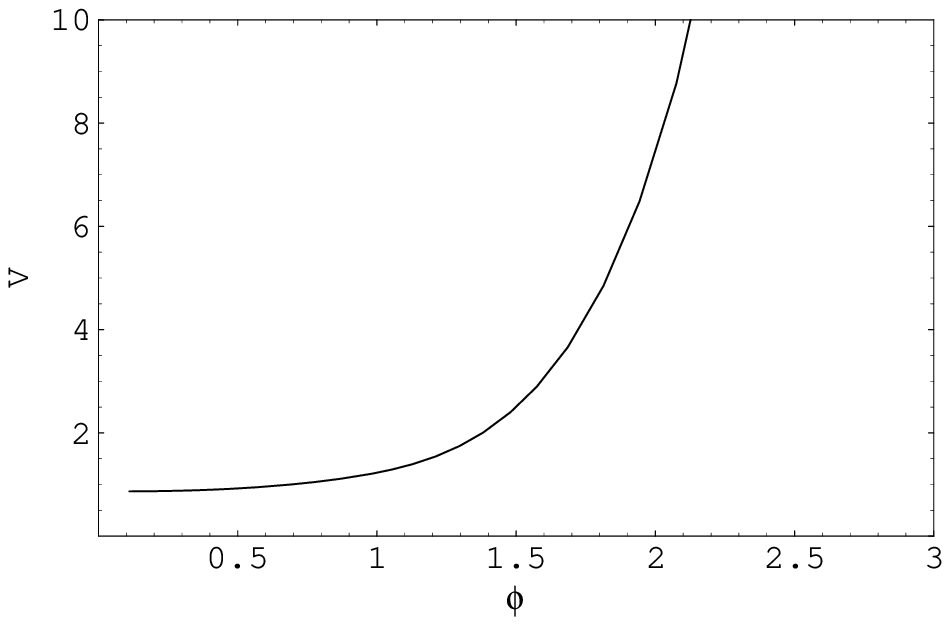}\\
Fig.3

\vspace{7mm} Figs. 1 - 3 shows variation of $\phi$ and $V$ against
$a$ and $\phi$ for $A(=\gamma)=1/3$ and $\alpha= 1$ (values of
other constants: $B=1, C=1, d=1$).\hspace{12.5cm} \vspace{6mm}

\end{figure}

Since models trying to provide a description of the cosmic
acceleration are proliferating, there exists the problem of
discriminating between the various contenders. To this aim Sahni
et al [15] proposed a pair of parameters $\{r,s\}$, called {\it
statefinder} parameters. In fact trajectories in the $\{r,s\}$
plane corresponding to different cosmological models demonstrate
qualitatively different behaviour. The above statefinder
diagnostic pair has the following form:

\begin{equation}
r=\frac{\dddot{a}}{aH^{3}}~~~~\text{and}~~~~s=\frac{r-1}{3\left(q-\frac{1}{2}\right)}
\end{equation}

where $H\left(=\frac{\ddot{a}}{a}\right)$ and
$q~\left(=-\frac{a\ddot{a}}{\dot{a}^{2}}\right)$ are the Hubble
parameter and the deceleration parameter respectively. The new
feature of the statefinder is that it involves the third
derivative of the cosmological radius. These parameters are
dimensionless and allow us to characterize the properties of dark
energy. Trajectories in the $\{r,s\}$ plane corresponding to
different cosmological models, for example $\Lambda$CDM model
diagrams correspond to the fixed point $s=0,~r=1$.\\

For one fluid model, these $\{r,s\}$ can be written as

\begin{equation}
r=1+\frac{9}{2}\left(1+\frac{p}{\rho}\right)\frac{\partial
p}{\partial\rho}
\end{equation}
and
\begin{equation}
s=\left(1+\frac{\rho}{p}\right)\frac{\partial p}{\partial\rho}
\end{equation}

For the two component fluids, equations (13) and (14) take the
following form:

\begin{equation}
r=1+\frac{9}{2(\rho+\rho_{_{1}})}\left[\frac{\partial
p}{\partial\rho}(\rho+p)+\frac{\partial
p_{_{1}}}{\partial\rho_{_{1}}}(\rho_{_{1}}+p_{_{1}})\right]
\end{equation}
and
\begin{equation}
s=\frac{1}{(\rho+\rho_{_{1}})}\left[\frac{\partial
p}{\partial\rho}(\rho+p)+\frac{\partial
p_{_{1}}}{\partial\rho_{_{1}}}(\rho_{_{1}}+p_{_{1}})\right]
\end{equation}

The deceleration parameter $q$ has the form:
\begin{equation}
q=-\frac{\ddot{a}}{aH^{2}}=\frac{1}{2}+\frac{3}{2}\left(\frac{p+p_{_{1}}}{\rho+\rho_{_{1}}}\right)
\end{equation}

For modified gas and barotropic equation states, we can set:

\begin{equation}
x=\frac{p}{\rho}=A-\frac{B}{\rho^{\alpha+1}}
\end{equation}
and
\begin{equation}
y=\frac{\rho_{_{1}}}{\rho}=\frac{\frac{d}{a^{3(1+\gamma)}}}{\left[\frac{B}{1+A}+\frac{C}{a^{3(1+A)(1+\alpha)}}
\right]^{\frac{1}{1+\alpha}}}
\end{equation}

Thus equations (15) and (16) can be written as

\begin{equation}
r=1+\frac{9s}{2}\left(\frac{x+\gamma y}{1+y}\right)
\end{equation}
and
\begin{equation}
s=\frac{(1+x)\{A(1+\alpha)-\alpha x \}+\gamma(1+\gamma)y}{x+\gamma
y}
\end{equation}
with
\begin{equation}
y=\left[\frac{d^{(1+\alpha)(1+A)}B^{\gamma-A}(1+\gamma)^{1+\gamma}
} {C^{1+\gamma}(1+A)^{1+\gamma}(A-x)^{\gamma-A} }
\right]^{\frac{1}{(1+\alpha)(1+A)}}
\end{equation}

\begin{figure}
\includegraphics[height=1.7in]{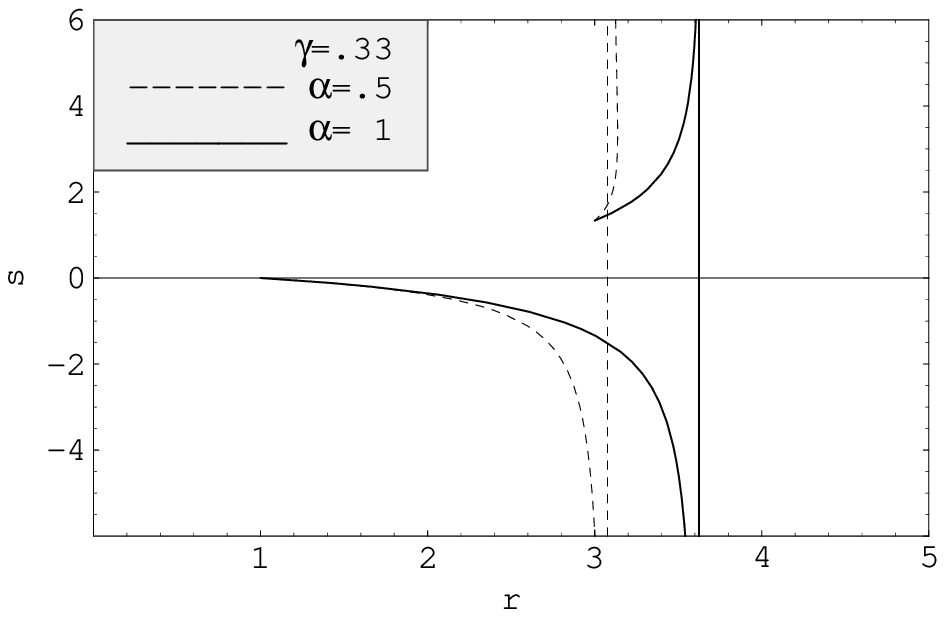}
\includegraphics[height=1.7in]{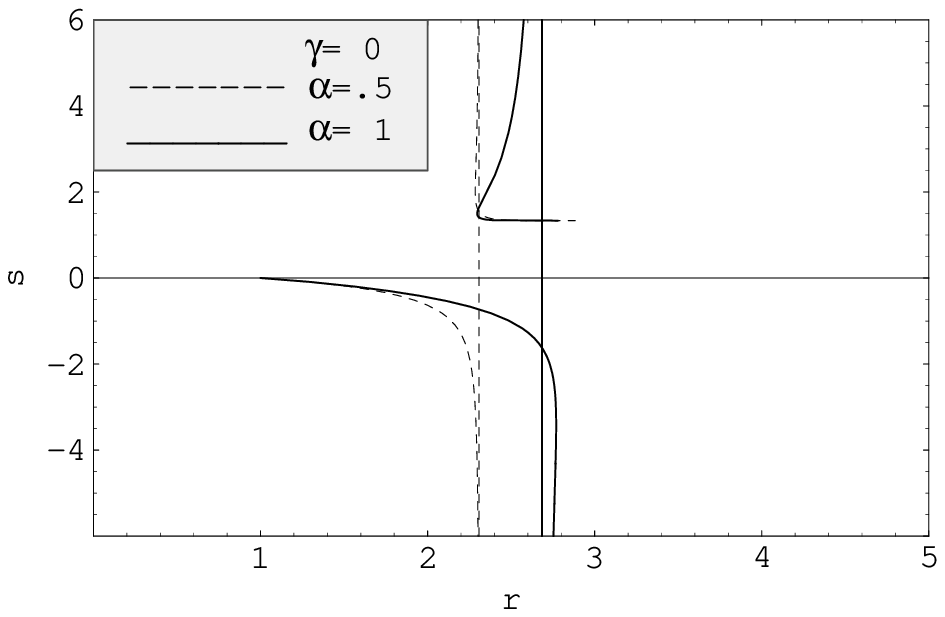}\\
\vspace{1mm}
Fig.4~~~~~~~~~~~~~~~~~~~~~~~~~~~~~~~~~~~~~~~~~~~~~~~~~~~~~Fig.5\\
\vspace{5mm}

\includegraphics[height=1.7in]{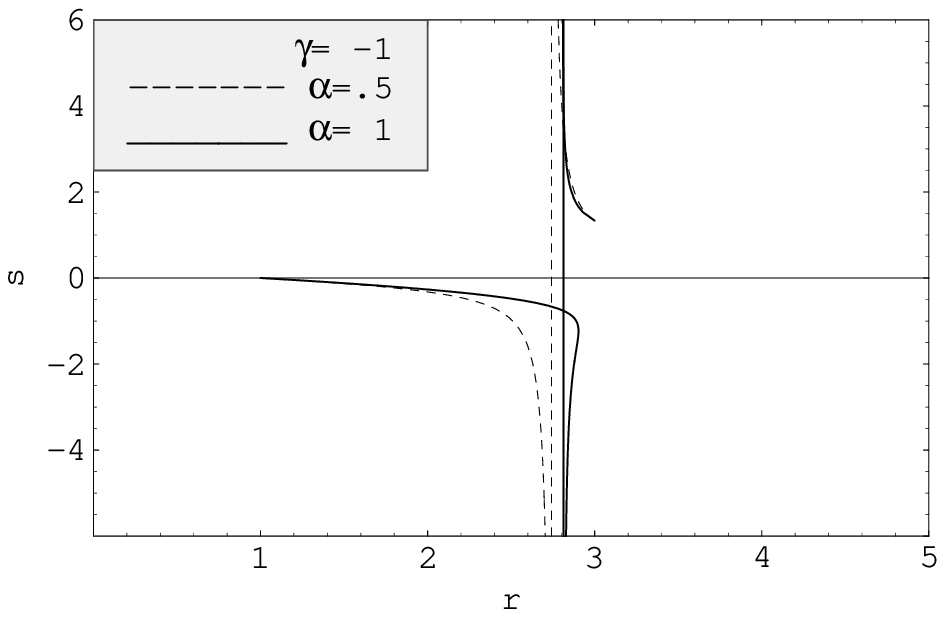}
\includegraphics[height=1.7in]{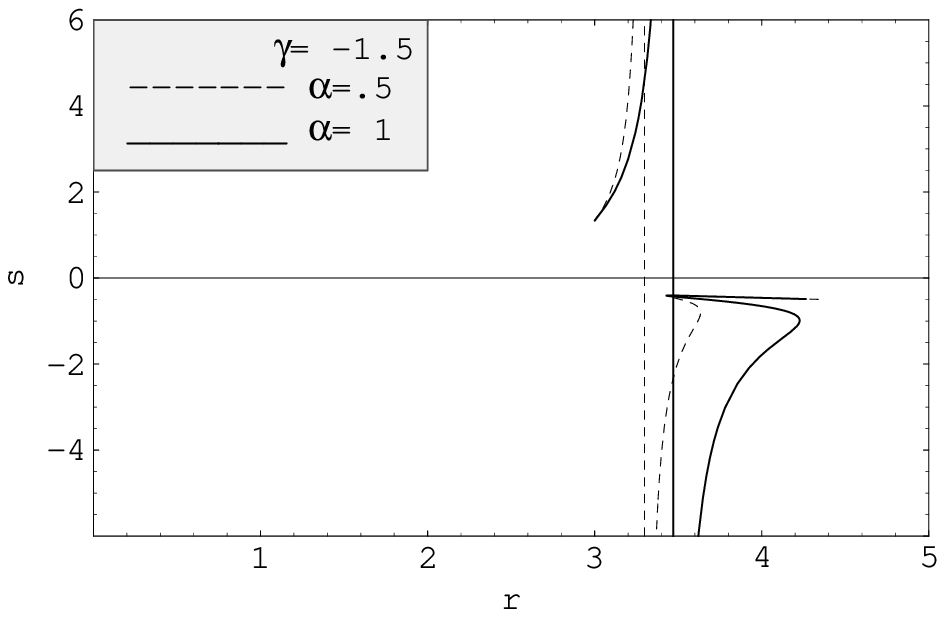}\\
\vspace{1mm}
Fig.6~~~~~~~~~~~~~~~~~~~~~~~~~~~~~~~~~~~~~~~~~~~~~~~~~~~~~Fig.7\\
\vspace{5mm}

\includegraphics[height=1.7in]{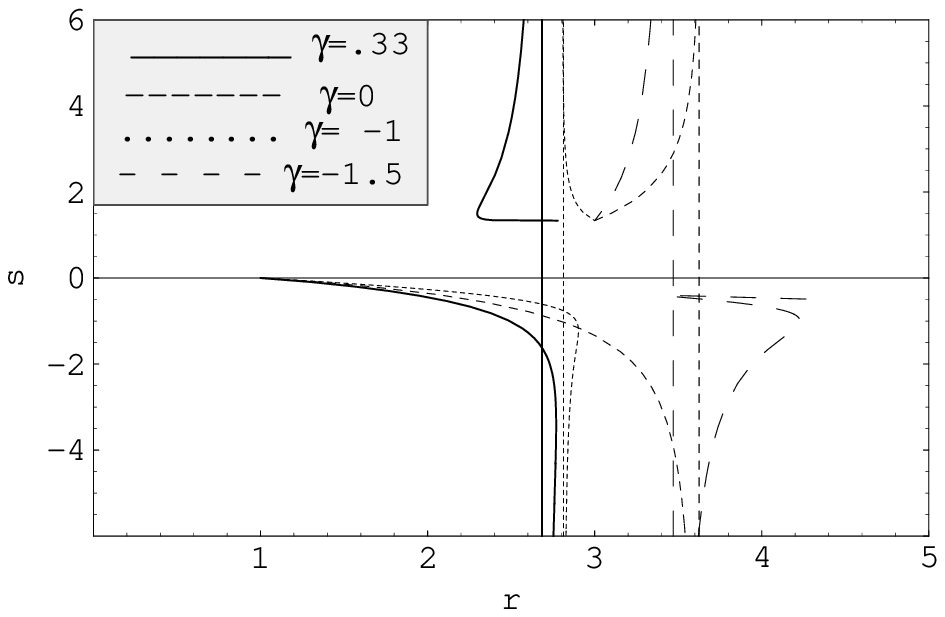}\\
\vspace{1mm}
Fig.8\\
\vspace{5mm}

\vspace{5mm} Fig. 4 - 7 show the variation of $s$
 against $r$ for different values of $\gamma=1/3,~0,~-1,~-1.5$ respectively and for
 $\alpha$ (= ~0.5,~1), $A=1/3$. Fig. 8 shows the variation of $s$ against $r$ for
 different values of $\gamma=1/3,~0,~-1,~-1.5$ and for
 $\alpha$ (=~1), $A=1/3$. \hspace{14cm} \vspace{4mm}

\end{figure}

Now for cosmic acceleration $q<0$, we have
$x+\gamma<-\frac{1}{3}$, since $y>0$. Here $y$ represents the
ratio of the energy density of the barotropic fluid to that of
Chaplygin gas and  $x, \gamma$ represent the ratios of fluid
pressure to energy density for barotropic fluid and Chaplygin gas
respectively. Since $\gamma$ is a  constant, we can assign
different values to this barotropic index. But for cosmic
acceleration we see that at least one of the fluids must generate
negative pressure. Also for $\gamma>\frac{2}{3}$ we must have
$x<-1$, which is not possible physically as the Chaplygin gas can
explain the evolution of the Universe only to $\Lambda$CDM model
[14]. Here for different values of $\gamma$ we have different
scenarios, viz, $(i)$ For $\gamma=\frac{1}{3}$, we have
$x<-\frac{2}{3}$, i.e., we have cosmic acceleration without
having the barotropic fluid to violate the energy condition. Here
the Chaplygin gas represents the dark energy and the barotropic
fluid represents the dark mater. $(ii)$ $\gamma=0$ and
$x<-\frac{1}{3}$ also give cosmic acceleration. Here the
Chaplygin gas violates the energy condition whereas the other
fluid represents dust. This particular choice can also represent
the present era, i.e., $q=-\frac{1}{2}$ provided $x<-\frac{2}{3}$.
Like the previous case here also the Chaplygin gas represents the
dark energy and the dust represents the dark matter. $(iii)$ For
$\gamma=-1$ we get cosmic acceleration without having the
Chaplygin gas to violate the energy condition. This model can
also represent the present epoch as for $q=-\frac{1}{2}$ and
$\gamma=-1$ we have $x<\frac{1}{3}$. Here the barotropic fluid
represents the dark energy and most likely behaves as the
cosmological constant and itself is enough to generate cosmic
acceleration. Here the Chaplygin gas represents the dark energy
or the dark matter according as the value of $x$. $(iv)$ For
$\gamma<-1$ this fluid represents the phantom model whereas the
Chaplygin gas represents the dark
energy or the dark matter according as the value of $x$. \\

The above four cases have been considered taking some particular
values of $\gamma$. We have discussed the possibility of both the
fluids to represent dark energy or dark matter. For
$\frac{2}{3}\geq \gamma>-\frac{1}{3}$ the Chaplygin gas represents
the dark energy and the barotropic fluid represents the dark
matter. For $-\frac{1}{3}\geq \gamma\geq -1$ the barotropic fluid
represents the dark energy. In this case the Chaplygin gas can
represent both dark energy or dark matter depending on the values
of the other parameters and the the ratio of th energy densities
of the two fluids. For $\gamma<-1$ the model represents phantom
energy.\\

From the equations (20) and (21) we can not written the
relationship between $r$ and $s$ in closed form. Thus the
relation between the parameters $r$ and $s$ in $\{r,s\}$ plane
for different choices of other parameters are plotted in figures
4 - 8. The figures 4 - 7 shows the variation of $s$ against $r$
for different values of $\gamma=1/3,~0,~-1,~-1.5$ respectively
and for $\alpha$ (= ~0.5,~1), $A=1/3$. Fig. 8 shows the variation
of $s$ against $r$ for different values of
$\gamma=1/3,~0,~-1,~-1.5$ and for $\alpha$ (=~1), $A=1/3$. Thus
the figures 4 - 6 represent the evolution of the universe
starting from the radiation era to the $\Lambda$CDM model for
$\gamma=1/3,~0$ and the figure 7 represents the evolution of the
universe starting from the radiation era to the quiessence model
for $\gamma=-1.5$. Thus $\gamma$ plays an active role for the
various stages of the evolution of the universe. If we choose the
arbitrary constant $d$ is equal to zero, we recover the model of
Modified Chaplygin gas [14]. If $A$ and the barotropic index
$\gamma$ are chosen to be zero, we get
back to the results of the works of Gorini et al [10].\\\\

{\bf Acknowledgement:}\\

One of the authors (W.C.) is grateful to Anirban Sarkar for
helping her in a few calculations.\\

{\bf References:}\\
\\
$[1]$ S. J. Perlmutter et al, {\it Bull. Am. Astron. Soc.} {\bf
29} 1351 (1997).\\
$[2]$ S. J. Perlmutter et al, {\it Astrophys. J.} {\bf 517} 565
(1999).\\
$[3]$ A. G. Riess et al, {\it Astron. J.} {\bf 116} 1009 (1998).\\
$[4]$ P. Garnavich et al, {\it Astrophys. J.} {\bf 493} L53
(1998).\\
$[5]$ B. P. Schmidt et al, {\it Astrophys. J.} {\bf 507} 46 (1998).\\
$[6]$ V. Sahni and A. A. Starobinsky, {\it Int. J. Mod. Phys. A}
{\bf 9} 373 (2000).\\
$[7]$ P. J. E. Peebles and B. Ratra, {\it Rev. Mod. Phys.} {\bf
75} 559 (2003).\\
$[8]$ T. Padmanabhan, {\it Phys. Rept.} {\bf 380} 235 (2003).\\
$[9]$ A. Kamenshchik, U. Moschella and V. Pasquier, {\it Phys.
Lett. B} {\bf 511} 265 (2001).\\
$[10]$ V. Gorini, A. Kamenshchik and U. Moschella, {\it Phys. Rev.
D} {\bf 67} 063509 (2003).\\
$[11]$ U. Alam, V. Sahni , T. D. Saini and A.A. Starobinsky, {\it
Mon. Not. Roy. Astron. Soc.} {\bf 344}, 1057 (2003).\\
$[12]$ M. C. Bento, O. Bertolami and A. A. Sen, {\it Phys. Rev. D}
{66} 043507 (2002).\\
$[13]$ H. B. Benaoum, {\it hep-th}/0205140.\\
$[14]$ U. Debnath, A. Banerjee and S. Chakraborty, {\it Class.
Quantum Grav.} {\bf 21} 5609 (2004).\\
$[15]$ V. Sahni, T. D. Saini, A. A. Starobinsky and U. Alam, {\it
JETP Lett.} {\bf 77} 201 (2003).\\

\end{document}